\documentclass{pasj00}
\Received{2012~November~14}
\Accepted{2013~January~11}
\SetRunningHead{T. Fujinaga et al.}{An X-ray counterpart of HESS~J1427--608 discovered with Suzaku}
\draft
\usepackage{times}

\begin{document}

\title{An X-ray counterpart of HESS~J1427--608 discovered with Suzaku}
\author{Takahisa~Fujinaga,\altaffilmark{1,2} Koji~Mori,\altaffilmark{3} Aya~Bamba,\altaffilmark{4} 
Shoichi~Kimura,\altaffilmark{3} Tadayasu~Dotani,\altaffilmark{2,1,5} 
Masanobu~Ozaki,\altaffilmark{2} Keiko~Matsuta,\altaffilmark{5,2} Gerd~P\"uhlhofer,\altaffilmark{6} Hideki~Uchiyama,\altaffilmark{7} 
Junko~S.~Hiraga,\altaffilmark{7} Hironori~Matsumoto,\altaffilmark{8} and Yukikatsu~Terada\altaffilmark{9}}
\altaffiltext{1}{Department of Physics, Tokyo Institute of Technology, 2-12-1 Ookayama, Meguro-ku, Tokyo 152-8550}
\altaffiltext{2} {Institute of Space and Astronautical Science, Japan Aerospace Exploration Agency, \\ 3-1-1 Yoshinodai, Chuo-ku, Sagamihara, Kanagawa 252-5210}
\email{fujinaga@astro.isas.jaxa.jp}
\altaffiltext{3}{Department of Applied Physics and Electronic Engineering, University of Miyazaki, 1-1 Gakuen Kibanadai-Nishi, Miyazaki, 889-2192}
\altaffiltext{4}{Department of Physics and Mathematics, Aoyama Gakuin University, 5-10-1 Fuchinobe, Chuo-ku, Sagamihara, Kanagawa 252-5258}
\altaffiltext{5}{Department of Space and Astronautical Science, The Graduate University for Advanced Studies, \\ 3-1-1 Yoshinodai, Chuo-ku, Sagamihara, Kanagawa 252-5210}
\altaffiltext{6}{Institut f\" ur Astronomie und Astrophysik, Universit\" at T\" ubingen, Sand 1, D 72076 T\" ubingen, Germany} 
\altaffiltext{7}{The University of Tokyo, 7-3-1 Hongo, Bunkyo-ku, Tokyo 113-0033}
\altaffiltext{8}{Nagoya University, Furo-cho, Chikusa-ku, Nagoya 464-8602}
\altaffiltext{9}{Saitama University, Shimo-Okubo 255, Sakura-ku, Saitama 338-8570}

\maketitle

\begin{abstract}
We report the discovery of an X-ray counterpart of the unidentified very high energy gamma-ray source HESS~J1427--608. 
In the sky field coincident with HESS~J1427--608, an extended source was found in the 2--8~keV band, and was designated as Suzaku~J1427--6051. 
Its X-ray radial profile has an extension of $\sigma=\timeform{0.9'} \pm \timeform{0.1'}$ if approximated by a Gaussian. 
The spectrum was well fitted by an absorbed power-law with $N_{\rm H}=(1.1\pm0.3) \times 10^{23}~{\rm cm}^{-2}$, $\Gamma=3.1^{+0.6}_{-0.5}$, and the unabsorbed flux $F_{\rm X}=(9^{+4}_{-2}) \times 10^{-13}~{\rm erg}~{\rm s}^{-1}~{\rm cm}^{-2}$ in the 2--10~keV band. 
Using XMM-Newton archive data, we found seven point sources in the Suzaku source region. 
However, because their total flux and absorbing column densities are more than an order of magnitude lower than those of Suzaku~J1427--6051, we consider that they are unrelated to the Suzaku source. 
Thus, Suzaku~J1427--6051 is considered to be a truly diffuse source and an X-ray counterpart of HESS~J1427--608. 
The possible nature of HESS~J1427--608 is discussed based on the observational properties.
\end{abstract}

\section{Introduction}
\label{sec:intro}

The origin of cosmic rays is an unsolved problem since the discovery in 1912.  
The spectrum of cosmic rays has a break at the knee energy ($10^{15.5}$~eV), and those below the knee energy are believed to have Galactic origin. 
Recently, the High Energy Stereoscopic System (H.E.S.S.) found more than 50 sources on the Galactic plane in the very high energy (VHE; $\gtsim 100$~GeV) gamma-ray band (\cite{aharonian2005}; \cite{aharonian2006}). 
Because VHE gamma-rays are produced either through inverse Compton scattering of low energy photons by relativistic electrons, or through the decay of pions produced by collisions of relativistic protons with interstellar medium, high energy particles are surely present in the VHE gamma-ray sources. 
If relativistic electrons are present, they may be traced through the observations of synchrotron X-ray emission. 
However, in spite of extensive X-ray followup observations, a non-negligible fraction of the Galactic VHE gamma-ray sources do not have an X-ray counterpart, and they are referred to as unidentified (unID) sources  (e.g., \cite{fujinaga2011}). 

Among identified Galactic VHE gamma-ray sources, pulsar wind nebulae (PWNe) represent the largest population \citep{hinton2009}. 
Accordingly, one would expect that PWNe are the first candidate of the origin of the unID VHE gamma-ray sources. 
However, identifying the VHE gamma-ray sources as PWNe is not straightforward. 
The VHE gamma-ray emission region sometimes shows significant spatial offset from the X-ray emission region or a pulsar (e.g., \cite{dejager2009,mattana2009}). 
Furthermore, the life time of X-ray emitting electrons through synchrotron emission may be shorter than that of VHE gamma-ray emitting electrons through Compton scattering. 
This means that X-ray emission from the PWNe can be much dimmer than the VHE gamma-rays if the PWNe are old. 
Finally, a pulsar is not always found in the vicinity of (candidate) PWNe. 

HESS~J1427--608 is one of the unID VHE gamma-ray sources located at $l=314\fdg409,b=-0\fdg145$ \citep{aharonian2008}, and is extended with $\sigma=0\fdg06$\footnote{Poster presentation by Komin, N., et al. in ``The X-ray Universe 2011'' held in June 27--30, 2011, at Berlin, Germany, http://xmm.esac.esa.int/external/xmm\_science/workshops/2011symposium/posters/Komin\_TopicE.pdf}.  
The flux is $F_{\rm TeV}=4.0~\times~10^{-12}~{\rm erg~s}^{-1}~{\rm cm}^{-2}$ in the 1--10~TeV band with a photon index of $\Gamma=2.16$ \citep{aharonian2008}.  
No plausible counterpart is listed in the Galactic supernova remnant (SNR) catalog \citep{green2009} or in the SIMBAD data base\footnote{http://simbad.u-strasbg.fr/simbad/} in the vicinity of HESS~J1427--608. 
We also looked for a possible counterpart in the ATNF pulsar catalog \citep{manchester2005}. 
However, no energetic pulsar was found within \timeform{100'} from HESS~J1427--608. 
Here, the definition of the energetic pulsar is those with $\dot{E}/d^{2} > 10^{33}$~erg~s$^{-1}$~kpc$^{-2}$, where $\dot{E}$ is the spin-down energy and $d$ is the distance to the source. 
This search radius of \timeform{100'} is large enough compared to offsets of radio pulsars to VHE sources among known PWN systems, and virtually excludes the possibility of associating HESS J1427--608 with any known energetic radio pulsar. 
Thus, HESS~J1427--608 seems to be one of the most ``unlikely'' PWNe among the VHE gamma-ray sources. 
In order to search for an X-ray counterpart, we observed HESS~J1427--608 with Suzaku, and also analyzed the XMM-Newton archive data. 

\section{Observation and Data Reduction}
\label{sec:szobs}

We observed the sky region including HESS~J1427--608 with Suzaku (\cite{mitsuda2007}) from 2010 January 13 through 16. 
The journal of the observation is listed in table~\ref{tab:obslog}. 
Suzaku is equipped with two types of detectors: four sets of X-ray Imaging Spectrometers (XIS0--XIS3: \cite{koyama2007}) and a non-imaging Hard X-ray Detector (HXD: \cite{takahashi2007}; \cite{kokubun2007}). 
XIS1 uses the back-illuminated (BI) CCD while the others the front-illuminated (FI) CCDs. 
Among the four sensors, XIS2 is not operational\footnote{JX-ISAS-SUZAKU-MEMO-2007-08; http://www.astro.isas.jaxa.jp/suzaku/doc/suzakumemo/suzakumemo-2007-08.pdf} and a part of XIS0 (corresponding to the off-source region in the latter analysis) is not usable due to its anomaly\footnote{JX-ISAS-SUZAKU-MEMO-2010-01; http://www.astro.isas.jaxa.jp/suzaku/doc/suzakumemo/suzakumemo-2010-01.pdf}. 
Hereafter, unless otherwise mentioned, we used only XIS1 and XIS3 data for the current analysis. 
XIS was operated in normal clocking mode without any window options. 
Spaced-row charge injection (\cite{prigozhin2008}; \cite{uchiyama2009a}) was used to reduce the effects of radiation damage.

We used version 2.4.12.27 of the processed data for HESS~J1427--608. 
The data was analyzed with the HEADAS software version 6.10 and XSPEC version 12.6.0. 
We used the cleaned event file created by the Suzaku team. 
The resultant effective exposure was 104~ks. 

\section{Analysis}
\label{sec:szanalysis}

\subsection{Images}
\label{subsec:xisimage}

Figure 1 shows XIS images in the soft (0.5--2~keV) and the hard (2--8~keV) bands. 
The images were corrected for exposure using an exposure map generated by the FTOOL {\tt xisexpmapgen} after subtracting non X-ray background (NXB) images generated by the FTOOL {\tt xisnxbgen} \citep{tawa2008}. 
Finally, the XIS1 and XIS3 images were added, and the resulting image was binned to a pixel size of \timeform{8"} and smoothed with a Gaussian of $\sigma=\timeform{1.0'}$. 
The position center and the extent of HESS~J1427--608 are shown as a black cross and a yellow dashed circle, respectively. 
A few faint, point-like sources are seen in the soft band, while an apparently extended source is detected in the hard band. 
The position and apparent extension of the hard source match those of HESS~J1427--608. 
As will be shown in a later section, 
all the Suzaku soft-band sources may be explained by the point-like sources in the XMM-Newton archive data 
while the Suzaku hard-band source seems to be a truly new source. 
We thus designate the central hard-band source Suzaku~J1427--6051, and mainly focus on this source in the following analysis. 

Because HESS~J1427--608 and Suzaku~J1427--6051 are both extended, we evaluate the X-ray size of the source. 
For this purpose, careful estimation of the X-ray background is necessary since Suzaku~J1427--6051 is relatively faint. 
In particular, this region contains significant contribution from the Galactic ridge X-ray emission (GRXE) as well as the cosmic X-ray background (CXB), both of which are subject to mirror vignetting effect resulting in a centrally-peaked spatial distribution. 
Therefore, it is not appropriate to assume a flat background image defined at an off-source region within the field of view (FOV). 
Instead, we estimated the GRXE+CXB image in the source region through Monte-Carlo simulation.  We explain details of the X-ray background estimation below. 

First, we estimated the surface brightness of the GRXE and CXB assuming their uniform distribution in the sky. 
For this purpose, we analyzed the X-ray spectrum in the off-source annulus region shown in figure~\ref{fig:xisspecreg}. 
The region (between \timeform{6'.5} and \timeform{7'.5} from the center of Suzaku~J1427--6051) was selected to minimize contamination from Suzaku~J1427--6051. 
In other words, the off-source region contains only the GRXE and the CXB after the subtraction of NXB. 
Figure~\ref{fig:xisanuspec} shows the spectra extracted from the off-source region. 
The spectra were modeled by optically thin thermal three temperature plasma with neutral iron emission line plus the CXB following \citet{uchiyama2009b}. 
In the course of model fitting, the hydrogen column density was fixed to $1.54 \times 10^{22}~{\rm cm}^{-2}$ determined by H{\sc i} observations (\cite{kalberla2005}), the photon index of the CXB to 1.4, and the surface brightness of the CXB to $5.4~\times~10^{-15}~{\rm erg}~{\rm s}^{-1}~{\rm cm}^{-2}~{\rm arcmin}^{-2}$ in the 2--10~keV band (\cite{kushino2002}). 
We made the ancillary response file using the FTOOL {\tt xissimarfgen} assuming a uniform emission in the sky. 
The model could reproduce the observed spectrum well. 
The best-fit parameters are listed in table~\ref{tab:xisanuspec}. 

Next, we simulated the GRXE+CXB event data assuming uniform emission with the best-fit spectrum model determined above using the FTOOL {\tt xissim} \citep{ishisaki2007}. 
In order to avoid an extra ambiguity due to poor statistics, the exposure time for the simulation was set to 500 ks. 
Then we extracted the GRXE+CXB spectrum in the source region and added it to the NXB in the source region to obtain the total background spectrum, following the instruction in section 5.5.2 of the Suzaku technical description\footnote{http://heasarc.gsfc.nasa.gov/docs/suzaku/prop\_tools/suzaku\_td/node8.html\#SECTION0085200000000000}. 
Figure~\ref{fig:xisspeccomp} shows the simulated background spectra of the XIS1 and XIS3 respectively as well as the source and off-source spectra. 
As can be seen from the spectra, the source region contains significant emission above the GRXE+CXB level. 

Figure~\ref{fig:xisprof} shows the radial profiles of the NXB-subtracted image and the simulated GRXE+CXB image in the 2--8~keV band. 
The difference is attributed to Suzaku~J1427--6051. 
Even if we consider the relatively broad half-power radius (\timeform{1.0'}; \cite{serlemitsos2007}) of the point spread function of Suzaku mirror, 
Suzaku~J1427--6051 is clearly extended. 
In order to estimate the source extent, we compared the observed radial profile with that of a model calculation for an extended source having a 2D-gaussian distribution. 
Instead of performing the $\chi^2$ fitting, we calculated $\chi^2$ for a set of $\sigma$, i.e. \timeform{0.5'}, \timeform{0.7'}, \timeform{0.8'}, \timeform{0.9'}, and \timeform{1.1'}. 
The simulated profiles were generated with {\tt xissim}. 
It turned out that none of them gave an acceptable fit; the radial profile of Suzaku~J1427--6051 has a core and a tail in comparison with the best-fit simulation data. 
We thus performed the $\chi^2$ test using only the core region ($r<\timeform{2.0'}$). 
We then could obtain an acceptable fit ($\chi^2=0.6$ for 3 degrees of freedom). 
The best-fit value obtained was $\sigma=\timeform{0.9'} \pm \timeform{0.1'}$ with 90\% confidence level. 

\subsection{Pulsations}
\label{subsec:xiscurve}

Although Suzaku~J1427--6051 is extended, we tried to search for pulsations in the XIS light curve as point sources may be hidden in the image. 
We extracted light curves of the source region in the 0.5--10~keV band and binned them to a time resolution of 16 sec.  
We calculated Fourier power spectra every 2048~bin ($\simeq$~9.1~hour) using the FTOOL {\tt powspec} to obtain the ensemble-averaged power spectrum density (PSD). 
The PSD of XIS1 is dominated by the orbital period of the satellite and its higher harmonics, probably due to the contamination of earth albedo. 
Thus we used only the PSD of XIS3 in the subsequent analysis. 
No significant peak was detected in the PSD with an upper limit of $\sim10$ (Poisson fluctuation level was normalized to 2.0), which corresponds to an rms amplitude of 0.0064~count~s$^{-1}$. 
Because the net count rate of the source is 0.021~count~s$^{-1}$, this corresponds to a relative amplitude of 31\%. 
Thus we conclude that no coherent pulsation is present in the light curves of Suzaku J1427--6051 with an upper limit of 31 \% between the pulse periods of 32~sec and 32768~sec. 

\subsection{Energy spectra}
\label{subsec:xisspec}

Figure~\ref{fig:xisspec} shows the background-subtracted energy spectra of Suzaku~J1427--6051. 
Here, the background spectrum (sum of GRXE, CXB, and NXB) was calculated with the method detailed in section~\ref{subsec:xisimage}. 
It is found that the spectrum of Suzaku~J1427--6051 is featureless and heavily absorbed. 
Although a hint of feature is seen in the 6--7~keV band, which may be due to the spacial variation of the iron emission line in GRXE, it is not statistically significant. 
We calculated the ancillary response file using the FTOOL {\tt xissimarfgen} for an extended source of a Gaussian profile with $\sigma=\timeform{0.9'}$. 
The spectrum was well modeled by either an absorbed power-law or a thermal model, whose best-fit parameters are listed in table~\ref{tab:xisparm}. 
However, a thermal origin is unlikely because the abundance is unreasonably small, and the X-rays are thus considered to be produced by non-thermal processes. 
The absorbed X-ray flux of the source is $3.1~\times~10^{-13}~{\rm erg}~{\rm s}^{-1}~{\rm cm}^{-2}$ in the 2--10~keV band. 

We also analyzed the spectrum of the HXD to search for emission of Suzaku~J1427--6051 above 10~keV. 
We used the background spectrum called ``bgd\_d'' provided by the Suzaku team \citep{fukazawa2009} and added the CXB to it using the FTOOL {\tt hxdpinnxbpi}. 
We used the XIS nominal-position response file categorized to epoch 6, released as a CALDB. 
We made the background-subtracted spectrum and fitted it with a power-law, whose photon index was fixed to $\Gamma=3.1$, as derived from the XIS analysis. 
The flux in the 15--40~keV band is ($1.6~\pm~1.1$) $\times~10^{-12}~{\rm erg}~{\rm s}^{-1}~{\rm cm}^{-2}$ ($\chi^2/{\rm d. o. f.}=35.21/27$) if no systematic error is taken into account. 
Based on \citet{fukazawa2009}, who showed the reproducibility of NXB in the 15--40~keV band is about 3\% in the 90\% confidence range, we constructed the NXB spectra with 3\% higher count rates to include the systematic error of the NXB reproduction. 
Then, a significant signal is no longer detected. 
We thus obtained the 90\% upper limit of $5.3 \times 10^{-12}~{\rm erg}~{\rm s}^{-1}~{\rm cm}^{-2}$ in the 15--40~keV band.

\section{Archival data of XMM-Newton}
\label{sec:xmmobs}

In order to estimate the contribution of point sources, we analyzed the archival data of XMM-Newton including the HESS~J1427--608 field. 
XMM-Newton (\cite{jansen2001}) carries three X-ray telescopes each equipped with a European Photon Imaging Camera (EPIC) at the foci. 
EPIC consists of two MOS CCD camera (MOS1 and MOS2: \cite{turner2001}) and a pn CCD camera (pn: \cite{struder2001}). 
The observation was carried out on 2007 August 9 for 24~ks. 
During the observation, EPIC was operated in full-frame mode with medium filter.

The data were analyzed with the Science Analysis Software (SAS) version 10.0.0, HEADAS version 6.10 and XSPEC version 12.6.0. 
Time intervals of enhanced background, which are caused by soft proton flares, were removed  with thresholds of 0.20~count~s$^{-1}$ for MOS1, 0.24~count~s$^{-1}$ for MOS2, and 0.6~count~s$^{-1}$ for pn, respectively, calculated for PATTERN=0 events above 10~keV. 
We used X-ray events of PATTERN=0--12 (for MOS cameras) and PATTERN=0--4 (for pn camera) for the image and spectral analyses. 
The resultant effective exposures were 21~ks (MOS1), 22~ks (MOS2), and 15~ks (pn), respectively. 
The journal of the observation is listed in table~\ref{tab:obslog}. 

\subsection{Images}
\label{subsec:epicimage}

Figure~\ref{fig:epicimage} shows the summed images of MOS1 and MOS2 in the soft (0.3--2~keV) and the hard (2--12~keV) bands. 
The XIS images are also overlaid in green contours. 
Using the SAS tool {\tt edetect\_chain}, a total of 16 point sources were detected within \timeform{9'} from the center of Suzaku~J1427--6051. 
Eleven and seven sources were found in the soft and the hard bands, respectively, and two sources were commonly detected in both bands. 
Names and count rates of the point sources are listed in table~\ref{tab:xmmps}. 
Summing up the count rates of the point sources located within the Suzaku source region (X1 through X7), we obtained $(3.4\pm0.4)\times10^{-3}$~count~s$^{-1}$ in the 0.5--2~keV band. 
This may be converted to the XIS BI count rate of $(2.4\pm0.3)\times10^{-3}$~count~s$^{-1}$ assuming the power-law spectrum in table 2. 
Because this is comparable to the rate actually observed with Suzaku ($(3.0\pm0.2)\times10^{-3}$~count~s$^{-1}$), the XIS data may be explained by the sum of the XMM-Newton point sources in the soft band. 
On the other hand, if we carry out a similar calculation for the hard band, the XMM-Newton point sources would produce $7 \times 10^{-4}$~ count~s$^{-1}$ in XIS. 
Because this corresponds to only $\simeq 8$~\% of the actual XIS count rate, Suzaku~J1427--6051 is difficult to be explained by the sum of the sources and may be a truly diffuse source.

\subsection{Energy Spectra}
\label{subsec: xmmspec}

We calculated the energy spectra of  only X1 and of the summed spectra of the central seven point sources (X1 through X7). 
We made the response file and the ancillary response file using the SAS tool {\tt rmfgen} and {\tt arfgen}, respectively. 
We fitted these spectra with the model of an absorbed power-law. 
Because of the poor statistics, the photon index was fixed to 3.1, which was determined by the spectral analysis of Suzaku~J1427--6051. 
The best-fit parameters are listed in table~\ref{tab:epicparm}. 
Figure~\ref{fig:epicspec} shows the summed spectra of the seven point sources. 
The absorbed X-ray flux of the summed spectra was $1.7~\times~10^{-14}~{\rm erg}~{\rm s}^{-1}~{\rm cm}^{-2}$ in the 2--10~keV band, which is about 5\% of the flux derived with Suzaku for Suzaku~J1427--6051. 
It is also found that the column density of the summed spectra is two orders of magnitude smaller than that derived for Suzaku~J1427--6051. 
These results confirm that Suzaku~J1427--6051 cannot be explained by the XMM-Newton point sources. 

\section{Discussion}
\label{sec:discuss}

\subsection{X-ray and Multi-Wavelength View of HESS~J1427--608}
\label{subsec:szvsxmm}

We discovered an apparently-diffuse hard X-ray source, Suzaku~J1427--6051 with Suzaku, and several point sources with XMM-Newton in spatial coincidence with HESS~J1427--608. 
We found that Suzaku~J1427--6051 could not be explained by the sum of the point sources detected with XMM-Newton, and concluded that it is a truly diffuse source. 
Even if time variability of the detected point-like sources is considered, it is unlikely that Suzaku~J1427--6051 is explained by them because the hydrogen column densities are quite different. 
The XMM-Newton point sources are most likely foreground sources. 
To explain Suzaku~J1427--6051 by the sum of several point sources would require that all those sources would be below detection threshold at the time of the XMM-Newton observation, which appears very unlikely. 
Thus, we concluded that Suzaku~J1427--6051 is an intrinsically diffuse source and the X-ray counterpart of HESS~J1427--608. 

\begin{figure*}
  \begin{center}
    \FigureFile(154mm,67mm){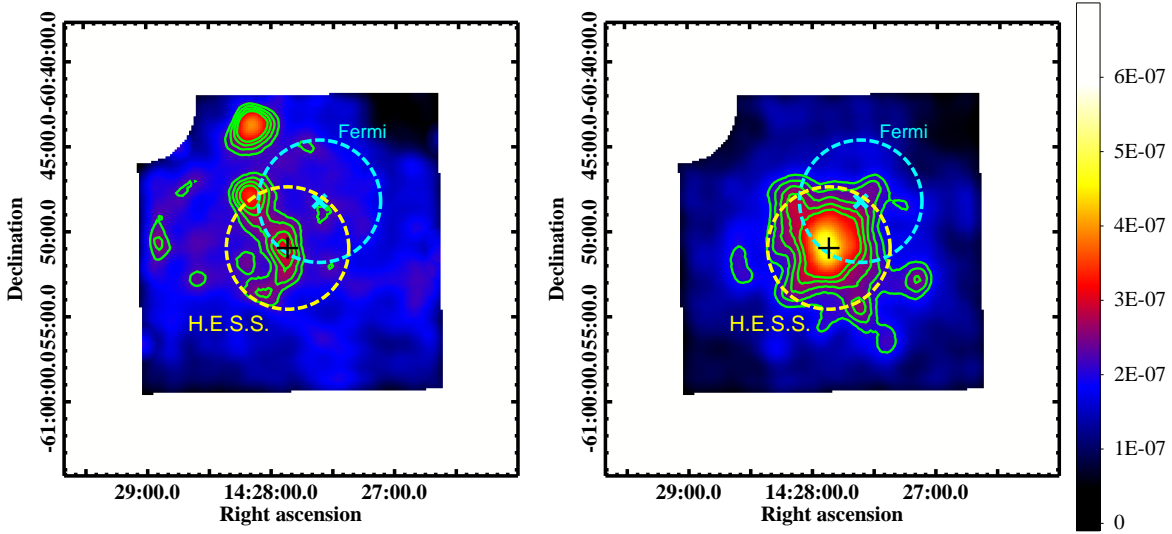}
  \end{center}
  \caption{XIS images of the HESS J1427--608 field in the 0.5--2~keV (left) and the 2--8~keV (right) bands, respectively. The color bar indicates the surface brightness in units of count s$^{-1}$ pixel$^{-1}$. The black cross and the dashed yellow circle indicate the position and the extent of the VHE source (\timeform{3.6'} in radius)\footnotemark[1], respectively. The cyan cross and the dashed cyan circle indicate the position and the error circle of 2FGL~J1427.6--6048c (\timeform{3.6'} in radius), respectively (\cite{nolan2012}). The green contours indicate the intensity of XIS image in linear scale. 
  }
  \label{fig:xisimage}
\end{figure*}

\begin{figure}
  \begin{center}
    \FigureFile(48mm,44mm){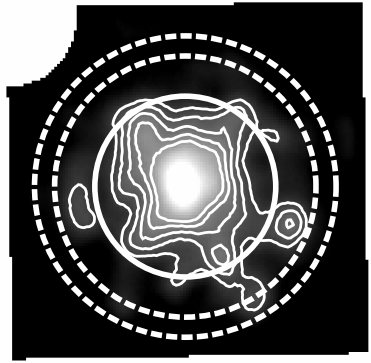}
  \end{center}
  \caption{The source and the off-source regions are indicated in the XIS hard band image. The contours shown in figure~\ref{fig:xisimage} are also overlaid. The source region is defined within the circle of $r=\timeform{4.5'}$ drawn by the solid line. The off-source region is defined as an annulus with an inner and an outer radius of $r=\timeform{6.5'}$ and $r=\timeform{7.5'}$ respectively, drawn by the dashed lines. }
  \label{fig:xisspecreg}
\end{figure}

\begin{figure}
  \begin{center}
    \FigureFile(80mm,60mm){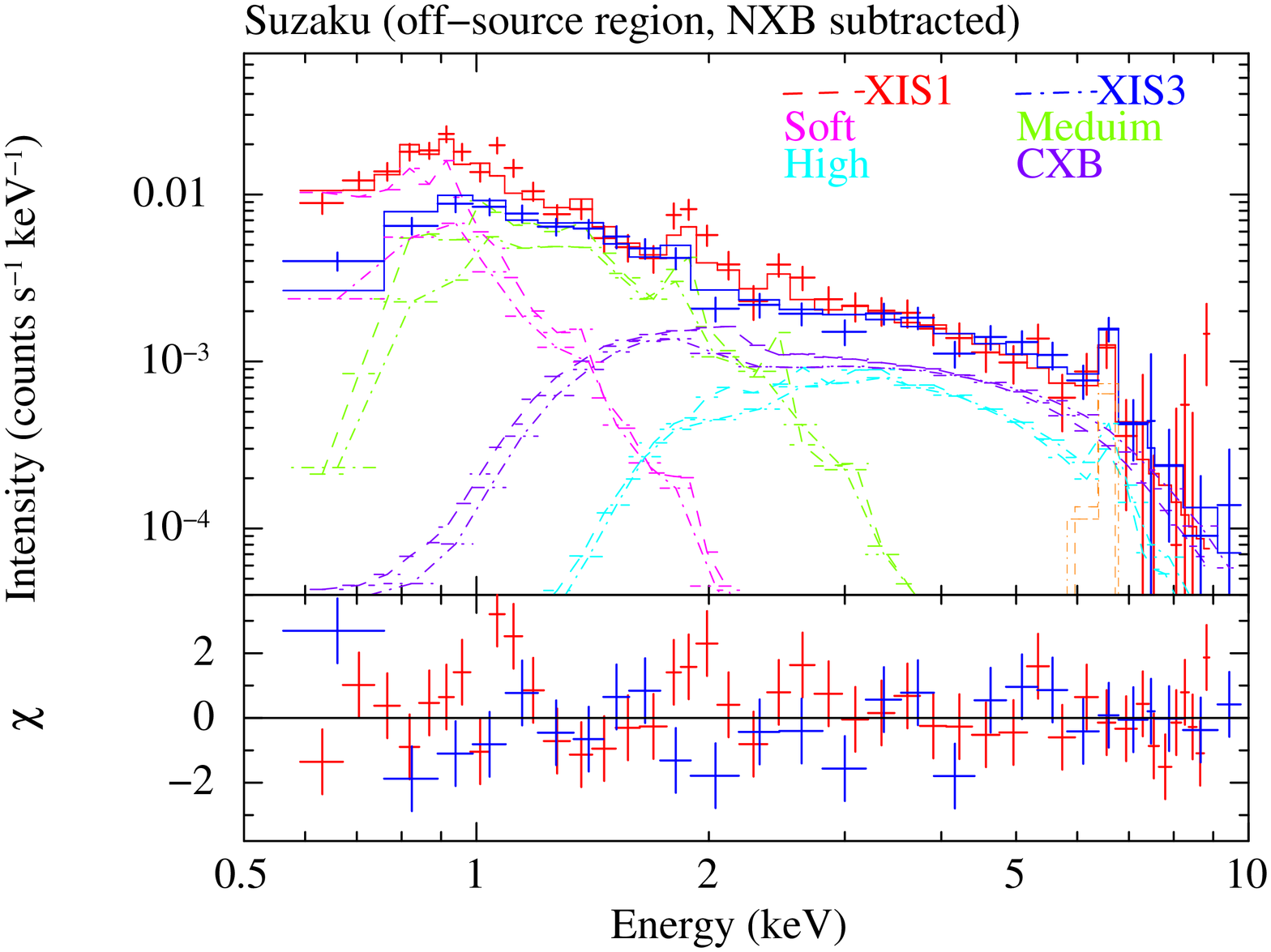}
  \end{center}
  \caption{XIS spectra of the off-source region after NXB subtraction. Red and blue data represent XIS1 and XIS3 spectra, respectively. The best-fit model of thin thermal three temperature plasma with an iron emission line plus the CXB is also shown. } 
  \label{fig:xisanuspec}
\end{figure}

\begin{figure}
  \begin{center}
    \FigureFile(80mm,60mm){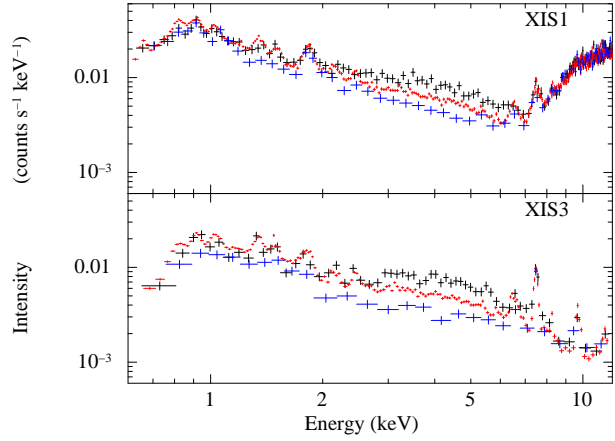}
  \end{center}
\caption{Comparison of the source spectrum (black) including the background, the background spectrum estimated for the source region (red), and the off-source spectrum (blue). The off-source spectrum is normalized considering the area difference of source and off-source regions. The upper and lower panels show spectra of XIS1 and XIS3, respectively. }
  \label{fig:xisspeccomp}
\end{figure}

\begin{figure}
  \begin{center}
    \FigureFile(80mm,60mm){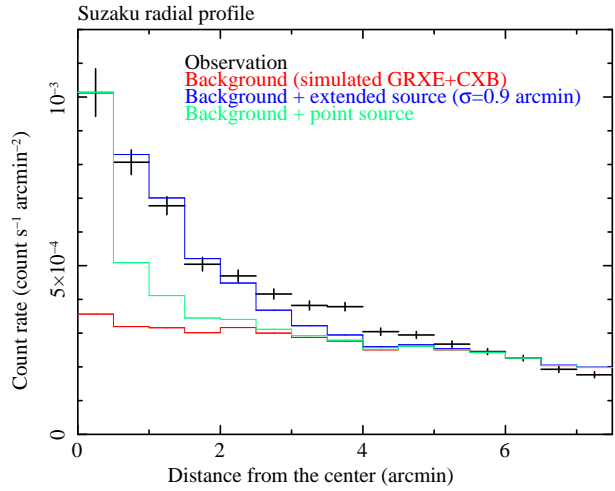}
  \end{center}
  \caption{The NXB subtracted, GRXE and CXB included radial profiles of XIS in the 2--8~keV band. Black, red, blue and green lines represent the observed radial profile, the simulated radial profile of GRXE+CXB, that of an extended source with $\sigma=\timeform{0.9'}$ and GRXE+CXB, and that of a point source and GRXE+CXB, respectively. }
  \label{fig:xisprof}
\end{figure}

\begin{figure}
  \begin{center}
    \FigureFile(80mm,60mm){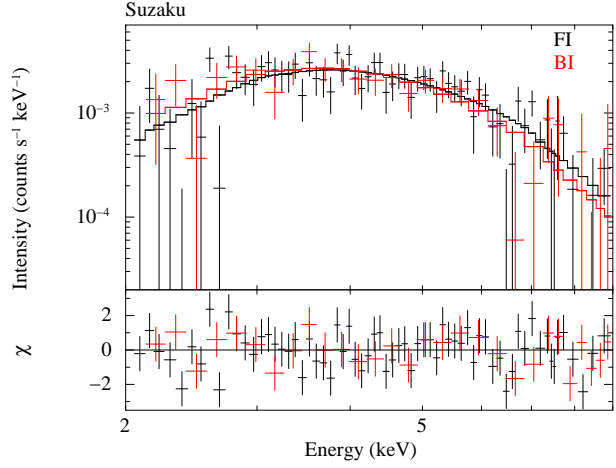}
  \end{center}
  \caption{The background-subtracted energy spectra of Suzaku~J1427--6051. Black and red crosses indicate data of FI CCDs (sum of XIS0 and XIS3) and BI CCD (XIS1), respectively. The solid lines show the best-fit absorbed power-law model. }
  \label{fig:xisspec}
\end{figure}

\begin{figure*}
  \begin{center}
    \FigureFile(130mm,59mm){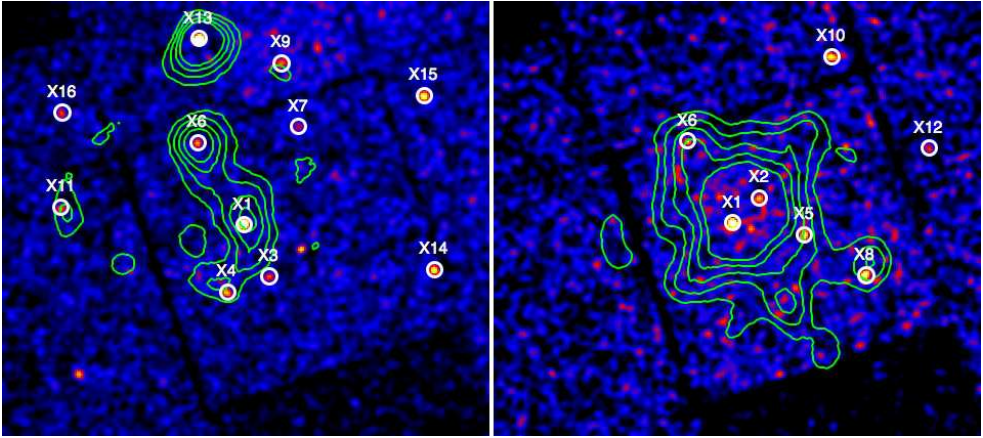}
  \end{center}
  \caption{The summed MOS images of the HESS J1427--608 field in the 0.3--2~keV band (left) and in the 2--12~keV band (right), respectively. The green contours indicate the XIS images shown in figure \ref{fig:xisimage}. The white circles indicate the point sources detected by the SAS tool {\tt edetect\_chain}, which are listed in table~\ref{tab:xmmps}. }
  \label{fig:epicimage}
 \end{figure*}

\begin{figure}
  \begin{center}
    \FigureFile(80mm,50mm){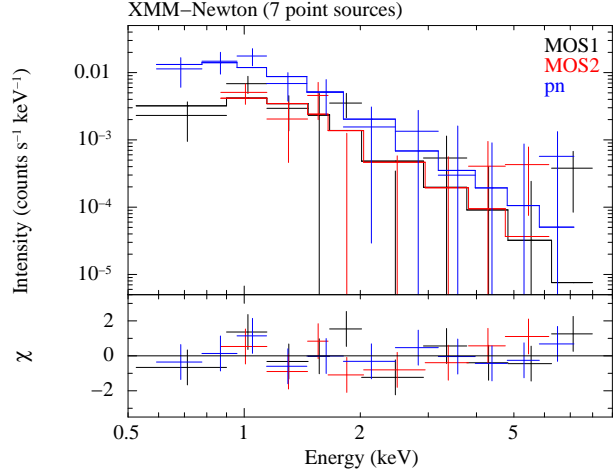}
  \end{center}
  \caption{The summed spectrum of seven point sources detected with XMM-Newton. Black, red and blue crosses indicate MOS1, MOS2 and pn spectra, respectively. The solid lines show the best-fit absorbed power-law model. }
  \label{fig:epicspec}
\end{figure}

We searched various catalogs and literature for the possible counterpart of HESS~J1427--608 in other wavebands. 
A GeV gamma-ray source, 2FGL~J1427.6--6048c, is listed in the 2-year catalog of Fermi (\cite{nolan2012}). 
The GeV source is located at $l=314\fdg3953, b=-0\fdg0909$, which is \timeform{3.3'} away from the center of HESS~J1427--608 (see figure~\ref{fig:xisimage}). 
Since the error radius of the position is about \timeform{3.6'}, it may be associated with HESS~J1427--608 \citep{nolan2012}. 
However, this is a ``c-designator-applied'' Fermi source whose position, emission characteristics, or even existence may not be reliable due to a potential confusion with interstellar emission \citep{nolan2012}. 
We believe it is premature to discuss a possible association of the GeV source to HESS~J1427--608; future Fermi data and analysis is necessary. 
We also searched the AKARI Point Source Catalogues (\cite{ishihara2010}; \cite{yamauchi2011}). 
No point source was found in the mid-infrared ($9~\mu$m) and the far-infrared ($90~\mu$m) bands within \timeform{1.5'} from the center of HESS~J1427--608.

\subsection{Possible Nature of HESS~J1427--608}
\label{subsec:relicpwn} 

\begin{figure}
  \begin{center}
    \FigureFile(80mm,50mm){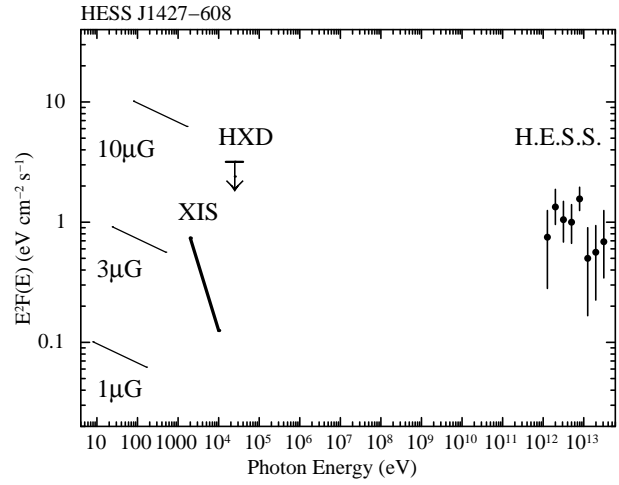}
  \end{center}
  \caption{Spectral energy distribution of HESS~J1427--608 from X-ray through VHE gamma-ray is shown. The calculated spectra of the synchrotron emission are overlaid. The H.E.S.S. spectrum is taken from \citet{aharonian2008}.}
  \label{fig:wideband}
\end{figure}

Although we mentioned HESS~J1427--608 as one of the most ``unlikely'' PWNe in the introduction, its center-filled morphology and featureless spectrum in the X-ray band suggest that the source could be a PWN. 
Thus we postulate here that the source is a PWN as a working hypothesis. 
Figure~\ref{fig:wideband} shows the spectral energy distribution (SED) from X-rays to VHE gamma-rays. 
The estimated synchrotron spectra from the VHE gamma-ray spectrum \citep{aharonian2008}, assuming that the same electron population would inverse-Compton scatter CMB photons up to VHE gamma-rays and radiate synchrotron emission in X-rays, are also plotted with a local magnetic field $B$ of 1, 3, and 10$~\mu{\rm G}$. 
This SED plot indicates that the simple one-zone leptonic model with $B$ of about 5$~\mu$G would roughly explain both the X-ray and VHE gamma-ray data. 
In this context, the steep Suzaku spectrum of $\Gamma \simeq 3.1$ could indicate that the Suzaku energy band is higher than the cut-off energy. 
The inferred magnetic field strength of $\simeq$5$~\mu$G is within the range of typical values on the Galactic plane. 
Thus the SED is consistent with a one-zone leptonic model expected for PWNe. 
We next examine the flux ratio and luminosities. 
The X-ray to VHE gamma-ray ratio is useful to probe the nature of unID sources \citep{yamazaki2006}. 
In the case of HESS~J1427--608, the unabsorbed X-ray flux $F_{\rm X}$ of $8.9 \times 10^{-13}~{\rm erg}~{\rm s}^{-1}~{\rm cm}^{-2}$ (2--10~keV) and the VHE gamma-ray flux $F_{\rm TeV}$ of $4.0~\times~10^{-12}~{\rm erg~s}^{-1}~{\rm cm}^{-2}$ (1--10 TeV) result in a flux ratio of $F_{\rm TeV}/F_{\rm X} \simeq~4.5$. 
If we assume a distance to HESS~J1427--608 of $d=8$~kpc, the X-ray luminosity would be $L_{\rm X} = 7 \times 10^{33}~{\rm erg}~{\rm s}^{-1}$. 
The estimated flux ratio and X-ray luminosity are within the values of known X-ray and VHE gamma-ray emitting PWNe, given that both values show a large scatter: $10^{-3} - 10^{2}$ in $F_{\rm TeV}/F_{\rm X}$ and $10^{32} - 10^{37}$~erg~s$^{-1}$ in $L_{\rm X}$ \citep{mattana2009}. 
Thus the flux ratio and luminosity are also consistent with sources of PWN origin. 

While there are observational evidences supporting the PWN origin as discussed above, others challenge such a view. 
With the assumption of $d=8$~kpc, the core size and whole radial extent would be 2~pc and 12~pc, respectively. 
The core size of 2~pc is a little large, but not exceptional among other PWN \citep{kargaltsev2008}.
The whole extent of 12~pc is not surprisingly large compared with the nebula of e.g., PSR~J1826--1334 \citep{uchiyama2009b}. 
The largest drawback of the PWN hypothesis is lacking the detection of both a pulsar and radio PWN associated. 
It is known that the pulsar luminosity and its nebula luminosity in the X-ray band are correlated over 7 orders of magnitude \citep{kargaltsev2008}. 
According to the correlation, the not-yet-detected pulsar should be brighter than $\sim 10^{-13}~{\rm erg}~{\rm s}^{-1}~{\rm cm}^{-2}$ in X-rays, one-tenth of the PWN luminosity. 
Such a pulsar would have been easily detected with XMM-Newton. 
In addition, the X-ray photon index of the nebula ($\Gamma = 3.1$) is very steep among the known X-ray nebulae most of which have a $\Gamma$ between 1.2--2.2 \citep{kargaltsev2008}.

In terms of photon index comparison, non-thermal SNRs might present a more plausible scenario than PWNe. 
Most of non-thermal SNRs radiate synchrotron X-rays with a photon index of $\Gamma=2.4-3.1$ \citep{nakamura2009}. 
The X-ray luminosity and extent are also compatible to synchrotron SNRs \citep{nakamura2012}. 
However, the facts that the source has a center-filled morphology in the X-ray band and no detection of the shell structure in the radio band hamper the interpretation as a non-thermal SNR. 

In summary, we have not reached a firm conclusion on the nature of HESS~J1427--608. 
PWN and/or non-thermal SNR, which are not necessary exclusive to each other, are the prime candidates but a decisive observational evidence is lacking. 
Increasing statistics in Fermi data may reveal a pulsation from 2FGL~J1427.6--6048c. 
A deep radio observation would also be helpful if it could detect pulsations or reveal the morphology of the source. 

\section{Conclusion}
\label{sec:summary}

We observed HESS~J1427--608 with Suzaku, which is combined with the archival data analysis of XMM-Newton, and obtained following results. 
\begin{itemize}
    \item We discovered an X-ray counterpart, Suzaku~J1427--6051, of HESS~J1427--608. It is intrinsically extended ($\sigma=\timeform{0.9'}\pm\timeform{0.1'}$) and has a spectral shape of an absorbed power-law with a photon index of $\Gamma=3.1^{+0.6}_{-0.5}$. 
    \item Several faint point sources were found in spatial coincidence with HESS~J1427--608 using XMM-Newton archival data in the soft X-ray band. 
    \item Observational properties of Suzaku~J1427--6051 and HESS~J1427--608 are compared with those of known PWNe and non-thermal SNRs. Some properties favor the PWN and/or SNR origin, but the available data are insufficient to draw firm conclusions. 
\end{itemize}

\bigskip

We thank Kentaro~Someya, Yoshitomo~Maeda and Michito~Sakai for their useful comments. 
T.F.\ acknowledges the financial support from the Global Center of Excellence Program by MEXT, Japan through the ``Nanoscience and Quantum Physics'' project of the Tokyo Institute of Technology, and JSPS young research fellowship, no.~23-9676. 
The work of K.M.\ is partially supported by the Grant-in-Aid for Young Scientists (B) of the MEXT (No.\ 24740167).

\begin{longtable}{*{5}{c}}
\caption{Journal of the observations}
\label{tab:obslog}
\hline
\hline
 & Obs ID & Observation date & Aim point\footnotemark[$*$] & Effective Exposure (ks) \\
\hline
\multicolumn{1}{l}{Suzaku} & 504034010  & 2010/01/13--2010/01/16 & ($314\fdg409, -0\fdg145$) & 104 \\
\hline 
\multicolumn{1}{l}{XMM-Newton} & 0504990101 & 2007/08/09 & ($314\fdg362, -0\fdg143$) & 21/22/15\footnotemark[$\dag$] \\
\hline
\multicolumn{5}{l}{\footnotemark[$*$] In Galactic coordinates.} \\
\multicolumn{5}{l}{\footnotemark[$\dag$] The effective exposures for MOS1/MOS2/pn. } \\
\endlastfoot
\end{longtable}

\begin{table}
 \caption{The best-fit parameters of the observed XIS spectrum in the annulus region}
 \label{tab:xisanuspec}
 \begin{center}
   \begin{tabular}{lccc}
   \hline 
   \hline 
 & Soft & Medium & High \\
   \hline
  $N_{\rm H}$ ($10^{22}$ cm$^{-2}$) & $0.21^{+0.17}_{-0.16}$ & $1.04^{+0.25}_{-0.19}$ & $4.95^{+10.86}_{-1.49}$ \\
  $kT$~(keV) & $0.26 \pm 0.06$ & $0.55^{+0.11}_{-0.08}$ & $3.71^{+11.45}_{-1.53}$ \\
  Absorbed flux\footnotemark[$*$] & $0.86^{+1.58}_{-0.66}$ & $3.30 \pm 1.06$ & $3.80^{+4.94}_{-1.90}$ \\
   &     \multicolumn{3}{c}{Abundance ($z_\odot$)} \\
   Ne & \multicolumn{3}{c}{0.57 (fixed)} \\
   S & \multicolumn{3}{c}{1.27 (fixed)} \\
   Ar & \multicolumn{3}{c}{2.10 (fixed)} \\
   others & \multicolumn{3}{c}{0.33 (fixed)} \\
  \hline 
   Line $E$ (keV) & \multicolumn{3}{c}{$6.47^{+0.22}_{-0.07}$} \\
   Line flux\footnotemark[$\dag$] & \multicolumn{3}{c}{$7.0^{+3.9}_{-3.5}$} \\
   \hline
   $\chi^2$/d.o.f & \multicolumn{3}{c}{84.68 / 60} \\
   \hline	
 \multicolumn{4}{l}{\footnotemark[$*$] In units of $10^{-12}~{\rm erg}~{\rm s}^{-1}~{\rm cm}^{-2}$ in the 0.8--10~keV band. }\\
 \multicolumn{4}{l}{\footnotemark[$\dag$] In units of $10^{-5}~{\rm photons}~{\rm s}^{-1}~{\rm cm}^{-2}$.} \\
   \end{tabular}
 \end{center}
\end{table}

\begin{table}
 \caption{The best-fit parameters of the XIS spectrum}
 \label{tab:xisparm}
 \begin{center}
   \begin{tabular}{lcc}
   \hline
   \hline
   & Power-law & APEC \\
   \hline 
  $N_{\rm H}$ ($10^{22}$~cm$^{-2}$) & $11.1^{+2.9}_{-2.5} $ & $9.0^{+2.2}_{-1.9}$ \\
  $\Gamma$ & $3.1 ^{+0.6}_{-0.5}$  & --- \\
  $kT$~(keV) & --- & $3.2^{+1.5}_{-0.8}$\\
  Abundance~($z_\odot$) & --- & $<0.09$\\
  Unabsorbed flux\footnotemark[$*$]  & $8.9^{+3.6}_{-2.0} $ & $6.6^{+4.7}_{-2.5}$ \\
   \hline
  Absorbed flux\footnotemark[$*$] & 3.1 & 3.0 \\
   \hline
    $\chi^2/{\rm d.o.f.}$ & $64.02 / 71$ & $62.89 / 70$ \\
   \hline
   \multicolumn{3}{l}{Note: The errors are in the 90~\% confidence range.} \\
   \multicolumn{3}{l}{\footnotemark[$*$] In the 2.0--10.0~keV band in units of $10^{-13}~{\rm erg}~{\rm s}^{-1}~{\rm cm}^{-2}$. } \\
   \end{tabular}
 \end{center}
\end{table}

\begin{longtable}{*{5}{c}}
\caption{Point sources detected with XMM-Newton}
\label{tab:xmmps}
\hline
\hline
ID & Name & Distance\footnotemark[$*$]  & \multicolumn{2}{c}{MOS Count rate\footnotemark[$\dag$]} \\
 & & & 0.3--2~keV & 2--12~keV \\
\hline
X1 & XMMU~J142755.4--605112.9 & \timeform{0.47'} & 5.3 & 3.3 \\
X2 & XMMU~J142746.5--605011.4 & \timeform{1.05'} & --- & 1.9 \\
X3 & XMMU~J142746.8--605317.8 &  \timeform{2.27'} & 2.3 & --- \\
X4 & XMMU~J142800.7--605356.4 & \timeform{3.13'} & 2.7 & --- \\
X5 & XMMU~J142731.4--605141.4 & \timeform{2.59'} & --- & 2.9 \\
X6 & XMMU~J142810.3--604752.0 & \timeform{3.85'} & 3.3 & 1.7 \\
X7 & XMMU~J142736.9--604712.8 & \timeform{4.21'} & 1.9 & --- \\
X8 & XMMU~J142711.0--605320.3 & \timeform{5.51'} & --- & 2.2 \\
X9 & XMMU~J142742.7--604436.9 & \timeform{6.48'} & 3.4 & --- \\
X10 & XMMU~J142722.3--604425.8 & \timeform{7.50'} & --- & 3.1 \\
X11 & XMMU~J142856.1--605027.9 & \timeform{7.83'} & 3.0 & --- \\
X12 & XMMU~J142649.9--604810.0 & \timeform{8.08'} & --- & 1.7 \\
X13 & XMMU~J142810.0--604337.9 & \timeform{7.69'} & 9.1 & --- \\
X14 & XMMU~J142651.3--605300.9 & \timeform{7.65'} & 4.9 & --- \\
X15 & XMMU~J142654.9--604556.1 & \timeform{8.60'} & 4.1 & --- \\
X16 & XMMU~J142855.6--604636.4 & \timeform{8.92'} & 3.1 & --- \\
\hline
\multicolumn{5}{l}{\footnotemark[$*$] Angular distance from the center of Suzaku~J1427--6051. } \\
\multicolumn{5}{l}{\footnotemark[$\dag$] Sum of MOS1 and MOS2 count rate in units of $10^{-3}$~count~s$^{-1}$. } \\
\endlastfoot
\end{longtable}

\begin{table}
\caption{The best-fit parameters of EPIC spectra}
\label{tab:epicparm}
\begin{center}
\begin{tabular}{lcc}
\hline
\hline
 & X1 & Sum of X1--X7 \\
\hline
$N_{\rm H}$~($10^{22}~{\rm cm}^{-2}$) & $<0.3$ & $0.3 \pm 0.1$ \\
$\Gamma$ & 3.1~(fixed) & 3.1~(fixed) \\
Unabsorbed flux\footnotemark[$*$] & $0.7^{+0.4}_{-0.2}$ & $1.7^{+0.7}_{-0.6}$ \\
\hline
Absorbed flux\footnotemark[$*$] & 0.7 & 1.7 \\
\hline
\multicolumn{1}{l}{$\chi^2/{\rm d.o.f.}$} & 11.28 / 6  &  16.74 / 27  \\
\hline
\multicolumn{3}{l}{Note: The errors are in the 90~\% confidence range.} \\
\multicolumn{3}{l}{\footnotemark[$*$]In the 2--10~keV band in units of $10^{-14}~{\rm erg}~{\rm s}^{-1}~{\rm cm}^{-2}$.} \\
\end{tabular}
\end{center}
\end{table}

\end{document}